\documentclass[prl,twocolumn,showpacs,superscriptaddress]{revtex4}

\usepackage{amsfonts}
\usepackage{amsmath}
\usepackage{amssymb}
\usepackage{amsthm}
\usepackage{bm}
\usepackage{dcolumn}
\usepackage{epsfig}
\usepackage{graphicx}
\usepackage{graphics}
\usepackage[latin1]{inputenc}
\usepackage{latexsym}
\usepackage{rotating}
\usepackage{hyperref}

\newcommand{\beqs}{\begin{equation*}}
\newcommand{\beq}{\begin{equation}}

\newcommand{\eeqs}{\end{equation*}}
\newcommand{\eeq}{\end{equation}}

\newcommand{\beqas}{\begin{eqnarray*}}
\newcommand{\beqa}{\begin{eqnarray}}

\newcommand{\eeqas}{\end{eqnarray*}}
\newcommand{\eeqa}{\end{eqnarray}}

\newcommand{\eq}[2]{\begin{equation} #1 \label{#2} \end{equation}}

\newcommand{\al}{\alpha}
\newcommand{\be}{\beta}
\newcommand{\ga}{\gamma}
\newcommand{\de}{\delta}

\newcommand{\ka}{\kappa}

\newcommand{\La}{\Lambda}

\newcommand{\blist}{\begin{itemize}}

\newcommand{\elist}{\end{itemize}}

\providecommand{\href}[2]{#2}

\DeclareFontFamily{OT1}{rsfs}{}
\DeclareFontShape{OT1}{rsfs}{m}{n}{ <-7> rsfs5 <7-10> rsfs7 <10->rsfs10}{} 
\DeclareMathAlphabet{\mycal}{OT1}{rsfs}{m}{n}

\DeclareMathOperator{\extdm}{d}
\newcommand{\extd}{\extdm \!}

\begin{document}
\title{Model for gravity at large distances}

\author{Daniel Grumiller}
\affiliation{Institute for Theoretical Physics, Vienna University of Technology, Wiedner Hauptstr. 8-10/136, A-1040 Vienna, Austria, Europe}

\date{\today}

\preprint{TUW-10-19}

\begin{abstract}
We construct an effective model for gravity of a central object at large scales. 
To leading order in the large radius expansion we find a cosmological constant, a Rindler acceleration, a term that sets the physical scales and subleading terms. 
All these terms are expected from general relativity, except for the Rindler term. 
The latter leads to an anomalous acceleration in geodesics of test-particles.
\end{abstract}


\maketitle


Gravity at large distances poses some of the most difficult puzzles in contemporary gravitational physics. 
The cosmological constant problem \cite{Carroll:2000fy} and the nature of dark matter \cite{Bertone:2004pz} are the most prominent ones. 
At a somewhat smaller scale, both in terms of actual size and in terms of scientific credibility, there are fly-by anomalies \cite{Anderson:2008zza} and the Pioneer anomaly \cite{Anderson:1998jd}. 
The word ``anomalous'' here refers to the difference between the observed trajectory of a test-particle in the gravitational field of a central object and the calculated trajectory. 
The pair test-particle/central object can mean e.g.~galaxy/cluster, star/galaxy, Sun/Pioneer spacecraft, Earth/satellite, etc.

Conceptually, there are three ways to resolve the anomalies. 
Either we modify the matter content of the theory (dark matter), 
or we modify the gravitational theory itself. 
The third alternative is that we might not be applying the theory correctly or beyond its realm of validity. 
Namely, even though effects that go beyond general relativity (GR) or its Newtonian limit are small locally, they may accumulate over large distances and/or through averaging, see \cite{Buchert:2002ij,Reuter:2004nx,Donoghue:2009mn} for various approaches that advocate this idea and implement it in different ways. 
In this Letter we advocate a new approach to describe gravity at large distances that is agnostic about the issue which of these three alternatives is realized in Nature.
The main strength of our method lies in its rigidity --- there is only one new free parameter --- and in its {\em ab initio} nature. 

One well-established key ingredient to our approach is to construct an effective field theory by writing down the most general action consistent with all the required symmetries and with additional assumptions (power counting renormalizability, analyticity, ...). 
A crucial point is that we impose spherical symmetry in addition to diffeomorphism invariance, which often is a good approximation in the IR~\footnote{We use the particle physicists' jargon and refer to the large distance limit as `infrared', abbreviated by IR.}. 
Spherical symmetry effectively reduces the theory to two dimensions. 
The main result of this Letter is that the most general action consistent with our assumptions leads to an additional term in the gravitational potential as compared to GR. This novel term generates accelerations similar to the ones observed in various ``anomalous'' systems in Nature and has a beautifully simple geometric interpretation as Rindler acceleration.

\section{Effective field theory for IR gravity}

To address the issue of anomalous acceleration of test-particles in the gravitational field of a central object we first simplify the underlying theory as much as possible. We assume that spacetime is described by a spherically symmetric metric in four dimensions (see e.g.~\cite{waldgeneral})
\eq{
\extd s^2=g_{\al\be}\,\extd x^\al\extd x^\be+\Phi^2\,\big(\extd\theta^2+\sin^2\theta\,\extd\phi^2\big)
}{eq:rind1}
The 2-dimensional metric $g_{\al\be}(x^\ga)$ and the surface radius $\Phi(x^\ga)$ depend only on the coordinates $x^\ga=\{t,r\}$. Our task is now clear: we should describe the dynamics of the fields $g_{\al\be}$ and $\Phi$. It is possible to do this in two dimensions, since the metric $g$ and the scalar field $\Phi$ are both intrinsically 2-dimensional objects. Each solution of the equations of motion (EOM) of that 2-dimensional theory leads to a 4-dimensional line-element \eqref{eq:rind1}, so the former faithfully captures the classical dynamics of the latter. The test-particles are then assumed to move in the background of the 4-dimensional line-element \eqref{eq:rind1}. The process of ``spherical reduction'' \cite{Berger:1972pg} simplifies the 4-dimensional Einstein--Hilbert action to a specific 2-dimensional dilaton gravity model, see e.g.~\cite{Grumiller:2002nm}. The novelty of our approach is that we allow for IR modifications of the dilaton gravity model. We explain now in detail how this works.

We construct the most general 2-dimensional theory with the field content $g$ and $\Phi$ compatible with the following assumptions. First of all, we require the theory to be power-counting renormalizable, assuming that non-renormalizable terms are suppressed. This leads uniquely to the action \cite{Russo:1992yg,Odintsov:1991qu}
\eq{
S=-\frac{1}{\ka^2}\,\int\!\extd^2x\sqrt{-g}\,\Big[f(\Phi)R+2(\partial\Phi)^2-2V(\Phi)\Big]
}{eq:rind2}
We have exploited field redefinitions to bring the kinetic term for the dilaton field $\Phi$ into a convenient form. 
The gravitational coupling constant $\ka$ does not play any essential role in our discussions. 
We assume that the free functions $f, V$ are analytic in $\Phi$ in the limit of large $\Phi$, like in spherically reduced GR \cite{Grumiller:2002nm}.
An analysis of the EOM (see below) reveals that the coupling function $f$ that multiplies the Ricci scalar $R$ must be given by $f=\Phi^2$ \footnote{Actually, even $f=\Phi^2+A\Phi+B+\dots$ is allowed. However, the $A\Phi$ term can be eliminated by a constant shift of $\Phi$, the $B$ term does not contribute to the EOM and the subleading terms can be neglected in the IR.} 
to reproduce the Newton potential $-M/r$. 
If we considered instead $f=c\,\Phi^2$ then the potential would change to $-M/r^{1/c}$. 
An experimental bound on $c$ is $|c-1|<10^{-10}$ \cite{Talmadge:1988qz}. 
Changing the power of $\Phi$ in the function $f$ would be even more drastic, leading to exponential growth or decay of the potential. 
In this work we make the conservative assumption that $f=\Phi^2$ is not renormalized in the IR, in excellent agreement with the experimental data.

We still have to choose the potential $V$ in the action \eqref{eq:rind2}. 
In the 4-dimensional language 
we consider large surface areas around some central object. 
After spherical reduction the limit of large surface areas leads to the limit of large dilaton field $\Phi$. 
We assume that the potential $V$ behaves as follows:
\eq{
V(\Phi)=\tilde\Lambda\Phi^2+\tilde a\Phi+\tilde b+{\cal O}(1/\Phi) 
}{eq:rind4}
The Ansatz \eqref{eq:rind4} for the potential $V$ is pivotal for the discussion, so let us pause and explain the rationale behind it. 
That the leading term in the large $\Phi$-expansion is quadratic follows from physics: if we did allow for powers higher than $\Phi^2$ 
then the ensuing metric would have a curvature singularity for large $\Phi$. 
Thus, \eqref{eq:rind4} is the generic asymptotic result provided we require the absence of curvature singularities and analyticity in $\Phi$ in the IR, which is what we do. 
The 
term ${\cal O}(1/\Phi)$ leads to a contribution to the gravitational potential proportional to $\ln{r}/r$, 
whose effects are subleading in the deep IR as compared to effects coming from the first three terms in 
\eqref{eq:rind4}. Therefore, we ignore this term (and further subleading terms). \footnote{%
The $\ln{r}/r$-term in the gravitational potential has a fall-off behavior comparable to the Newton term, but is logarithmically larger.
Thus, for comparison with precision data it may be worthwhile to take it into account, even though it is suppressed by a factor $r^2$ as compared to the Rindler term.
The next subleading term in the gravitational potential is a Reissner--Nordstr\"om term, which can safely be neglected in the IR.
Further subleading terms correspond not to IR but rather to UV modifications.
} 

By rescaling simultaneously $\Phi$ and $\ka$ we can fix one of the subleading coefficients in the asymptotic potential \eqref{eq:rind4}, which amounts to fixing a specific physical length scale. Without loss of generality we choose
$\tilde b \rightarrow b=-2$.
Dropping all asymptotically subleading terms and choosing convenient normalizations of the coupling constants as well as $\ka=1$ the action \eqref{eq:rind2} simplifies to
\eq{
S = - 
\int\!\extd^2x\sqrt{-g}\,\Big[\Phi^2R+2(\partial\Phi)^2-6\Lambda\Phi^2+8a\Phi+2\Big]
}{eq:rind6}
The action \eqref{eq:rind6} is our first key result. It provides the generic effective theory of gravity in the IR consistent with the assumptions spelled out above. The theory defined by the action \eqref{eq:rind6} depends on two coupling constants, $\Lambda$ and $a$. Solutions of the EOM descending from the action \eqref{eq:rind6} describe spherically symmetric line elements \eqref{eq:rind1} that model gravity in the IR. 

It is straightforward to find all solutions to the EOM 
\begin{align}
& R= \frac{2}{\Phi}\,g^{\al\be}\nabla_\al\partial_\be\Phi+6\La-\frac{4a}{\Phi}\\
& 2\Phi\,\big(\nabla_\mu\partial_\nu-g_{\mu\nu}\,\nabla^\al\partial_\al\big)\Phi-g_{\mu\nu}\,\big(\partial\Phi\big)^2=g_{\mu\nu}\,V(\Phi) 
\end{align}
derived from the action \eqref{eq:rind6}.
The result is~\footnote{This calculation is done most conveniently in the gauge theoretic formulation analog to \cite{Cangemi:1992bj}, see \cite{Grumiller:2002nm} for details. In addition to the generic solutions \eqref{eq:rind8}-\eqref{eq:rind9} there can be isolated degenerate solutions that we do not discuss here.}
\begin{align}
& 
g_{\al\be}\,\extd x^\al\extd x^\be = -K^2\extd t^2+\frac{\extd r^2}{K^2} \qquad \Phi = r
\label{eq:rind8} 
\\
& 
K^2 = 1-\frac{2M}{r}-\Lambda r^2+2ar
\label{eq:rind9}
\end{align}
The line-element \eqref{eq:rind8} exhibits a Killing vector $\partial_t$ with norm $K$ given by \eqref{eq:rind9}.
The 2-dimensional Ricci scalar $R=-\extd^2 K^2/\extd r^2 = 2\La+4M/r^3$ only depends on $\Lambda$ and on $M$, but not on $a$.
The quantity $M$ is a constant of motion. If $a=\Lambda=0$ we recover the Schwarzschild solution with $M$ being the black hole mass. 
If $\Lambda\neq 0$ we have additionally a cosmological constant.  
If $a\neq 0$ we obtain a contribution to the Killing norm $K$ that does not arise in vacuum Einstein gravity. 
The geometric meaning of this new term is clear \cite{Grumiller:2002nm}: it generates Rindler acceleration. 
Indeed, for $M=\La=0$ the line-element \eqref{eq:rind8} is the 2-dimensional Rindler metric \cite{waldgeneral}. 
Thus, our effective field theory \eqref{eq:rind6} that describes gravity in the IR differs in one aspect from GR: it allows for an arbitrary Rindler acceleration term. This is our second key result \footnote{There is a simple way to eliminate the Rindler term and recover GR with a cosmological constant $\La$, namely to require the action \eqref{eq:rind6} to be symmetric under $\Phi\to-\Phi$. However, there is no {\em a priori} reason for this symmetry.}.

\section{Physical consequences}

We take now the presence of the Rindler term in \eqref{eq:rind9} for granted and discuss some of its physical consequences.

\paragraph{Order of magnitude of the parameters} The cosmological constant is given by $\La\approx 10^{-123}$ \cite{Riess:1998cb,Perlmutter:1998np}, and this is the value we shall be using, though in several applications we set $\La=0$ for simplicity. Note that we are employing natural units $c=\hbar=G_N=1$. The Rindler acceleration $a$ in principle may depend on the scales of the system under consideration. We motivate now an order of magnitude estimate for $a$ in the deep IR. The product $\La r^2$ that enters the Killing norm \eqref{eq:rind9} becomes order of unity if $r$ is of the order of the radius of the visible Universe. If we suppose that the same happens for the $2a r$-term then we obtain the estimate $a\approx 10^{-62}-10^{-61}$, which is approximately the scale were anomalous accelerations are observed: the MOND acceleration is about $10^{-62}$ \cite{Milgrom:1983ca} and the Pioneer acceleration is about $10^{-61}$ \cite{Anderson:1998jd}.

\paragraph{Geodesics of test-particles} We study now geodesics of timelike test-particles (of unit mass) propagating on the 4-dimensional background determined by the spherically symmetric line element \eqref{eq:rind1} with \eqref{eq:rind8}-\eqref{eq:rind9}. Consider for simplicity motion in the plane $\theta=\pi/2$. Using standard methods \cite{waldgeneral} we obtain the well-known result $\dot\phi=\ell/r^2$, where $\ell$ is the conserved angular momentum, and 
\eq{
\frac{\dot r^2}{2}+ V^{\rm eff} = E
}{eq:rind10}
with $E=\rm const.$ The effective potential reads 
\eq{
V^{\rm eff} = -\frac{M}{r} + \frac{\ell^2}{2r^2} - \frac{M\ell^2}{r^3} - \frac{\La r^2}{2} + ar\,\big(1+\frac{\ell^2}{r^2}\big) 
}{eq:rind11}
The first term in $V^{\rm eff}$ is the Newton potential, the second the centrifugal barrier, the third the GR correction, the fourth a cosmic acceleration term and the last term proportional to the Rindler acceleration $a$ is novel. For vanishing cosmological constant and vanishing angular momentum the force $F^{\rm eff}$ derived from the effective potential contains only two terms.
\eq{
F^{\rm eff}=-\frac{\partial V^{\rm eff}}{\partial r}\Big|_{\ell=\La=0} = -\frac{M}{r^2}-a
}{eq:rind12}
Thus, for very small Newton potentials the Rindler acceleration term becomes relevant and provides a constant acceleration towards the source (if $a$ is positive) or away from the source (if $a$ is negative). This effect becomes important at sufficiently large distances only.

\begin{figure}
\centering
\epsfig{file=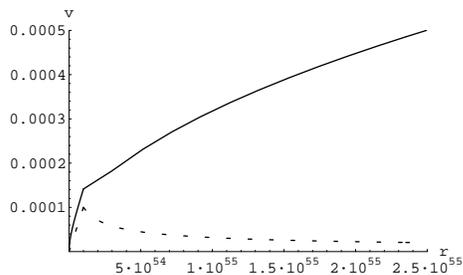,width=0.7\linewidth}
\caption{Toy model rotation curve for dwarf galaxy}
\label{fig:1}
\end{figure}

\begin{figure}
\centering
\epsfig{file=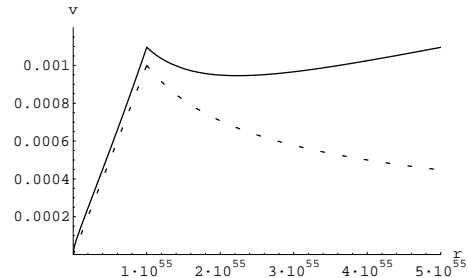,width=0.7\linewidth}
\caption{Toy model rotation curve for large spiral galaxy}
\label{fig:2}
\end{figure}

\paragraph{Galactic rotation curves} In the Newtonian limit the velocity profile is determined from the mass by $v(r)\approx \sqrt{M(r)/r}$, where we assume some radial mass profile $M(r)$ and set to zero the cosmological constant and the Rindler acceleration. If we take into account also the Rindler acceleration we obtain a new formula for the radial velocity profile in a galaxy:
\eq{
v(r)\approx \sqrt{\frac{M(r)}{r}+ar}
}{eq:rind13} 
Let us now construct a rough model of galaxies, starting with dwarf galaxies. We assume that the density is constant in the center up to $r\approx 10^{54}$ (less than a kiloparsec) and drops to zero thereafter. For a total mass of $M\approx 10^8$ solar masses we obtain from \eqref{eq:rind13} the velocity profile depicted in Fig.~\ref{fig:1}.
The dashed line gives the Keplerian velocity profile and the solid line gives the velocity profile deduced from \eqref{eq:rind13} with $a\approx 10^{-62}$. Strikingly, Fig.~\ref{fig:1} resembles the rotation curves for dwarf galaxies, see e.g.~\cite{Rhee:2003vw}. To describe large galaxies we assume that the density is constant in the center up to $r\approx 10^{55}$ (about 3 kiloparsecs) and drops to zero thereafter. For a total mass of $M\approx 10^{11}$ solar masses we obtain from \eqref{eq:rind13} the velocity profile depicted in Fig.~\ref{fig:2}. 
The dashed line gives the Keplerian velocity profile and the solid line gives the velocity profile deduced from \eqref{eq:rind13}, again with $a\approx 10^{-62}$. Strikingly, Fig.~\ref{fig:2} resembles the rotation curves for large (spiral) galaxies, see e.g.~\cite{Sofue:2000jx}. The scale where the velocity profile flattens is $v\approx 10^{-3}$, which roughly corresponds to $300 km/s$, in reasonable agreement with the data. 

It is worthwhile mentioning that the velocity profile \eqref{eq:rind13} was studied on purely phenomenological grounds, see Eq.~(125) in \cite{Mannheim:2005bfa} in section 6.3 (``Clues from data''). It is gratifying that the effective field theory \eqref{eq:rind6} predicts a velocity profile \eqref{eq:rind13} that was argued to be a good phenomenological fit to the data~\footnote{There is a slight difference to the phenomenological fit for the velocity profile studied in \cite{Mannheim:2005bfa}: while our Rindler acceleration $a$ is a prescribed constant, in Mannheim's work it consists of two terms, one of which is also a prescribed constant, while the other scales with the number of stars in the galaxy. The latter contribution does not arise in our scenario since we have assumed for simplicity that $a$ takes the same value in all galaxies. We can drop this assumption and recover precisely Mannheim's fit if we assume that $a$ depends on the scales of the system under consideration, $a=a_0+a_1 N$, where $N$ is the number of stars in the galaxy. 
For dwarf galaxies the contribution coming from $a_1$ does not play any role, while for large galaxies it plays only a modest role.}. 
We note, however, that so far there is no empirical hint that generically rotation curves of galaxies rise linearly at large distances, see e.g.~\cite{deBlok:1997ut}, so one has to take this rough model of galaxies with a few grains of salt. Of course, allowing for an $r$-dependent function $a$ in \eqref{eq:rind13} one can fit any given rotation curve. It could be a rewarding exercise to perform such fits to the data, thereby determining the Rindler acceleration $a$ as a function of $r$.

\paragraph{Solar system precision tests and Pioneer anomaly} Physics in the solar system appears to be well-understood. Nevertheless, there are a few tentative anomalies like fly-by anomalies and the Pioneer anomaly, for a review cf.~\cite{Lammerzahl:2006ex}. 
The force law \eqref{eq:rind12} leads to an anomalous constant acceleration towards the Sun (if $a$ is positive). Interestingly, this effect has been observed by the Pioneer spacecraft \cite{Anderson:1998jd}. Taking the data at face value requires to fix $a\approx 10^{-61}$, which is not that different from the value of $a$ used in the toy model for galaxies above. Even though there is still room for doubt if the Pioneer anomaly is a genuine effect \cite{Dittus:2005re}, it is striking that our effective field theory \eqref{eq:rind6} is capable to account for it.
It is notoriously difficult to reconcile modifications of gravitational theories that explain, for instance, the Pioneer anomaly without spoiling solar system precision tests \cite{Tangen:2006sa}. Our Rindler acceleration term does not necessarily spoil the solar-system precision tests, since the value of $a$ depends on the system that we describe. For instance, in the system Galaxy/Sun $a\approx 10^{-62}$, in the system Sun/Mercury $a$ may be much smaller, in the system Sun/Pioneer spacecraft $a\approx 10^{-61}$, and in the system Earth/satellite $a$ may be much bigger. It is a challenging open issue to understand precisely what determines the scale of $a$. It is unlikely that such an understanding can be achieved within our 2-dimensional effective model.


\paragraph{4-dimensional interpretation} The viewpoint taken so far was that the effective theory for gravity in the IR is a 2-dimensional scalar-tensor theory \eqref{eq:rind6}, whose solutions of the (vacuum) EOM lead to 4-dimensional spherically symmetric metrics \eqref{eq:rind1} with \eqref{eq:rind8}-\eqref{eq:rind9}. 
Following this philosophy we derived above some remarkable consequences. 
One issue, however, is not entirely satisfactory: we had to take a detour via spherical reduction to two dimensions to establish our results. 
It would be interesting to derive --- or at least to interpret --- the same results directly in four dimensions.
Such a top-down derivation is accessible only by reintroducing model assumptions that lift the agnosticism of the current approach.
In the present work we confine ourselves to possible 4-dimensional interpretations of the 2-dimensional Rindler acceleration.
Therefore, we pose now the question what kind of field equations are obeyed in four dimensions by metrics of type \eqref{eq:rind1} with \eqref{eq:rind8}-\eqref{eq:rind9}. Obviously, any metric can formally be written as a solution to Einstein's equations with some suitable energy momentum tensor. 
\eq{
R^\mu_\nu-\frac12\,\de^\mu_\nu\,R +3\,\de^\mu_\nu\, \La = 8\pi\,T^\mu_\nu
}{eq:rind14}
In our case this effective energy momentum tensor turns out to be like the one of an anisotropic fluid
\eq{
T^\mu_\nu = {\rm diag}\,(-\rho,\,p_r\,,p_\perp\,,p_\perp)^\mu_\nu
}{eq:rind15}
with density $\rho$, radial ($p_r$) and tangential ($p_\perp$) pressure
\eq{
\rho=-\frac{a}{2\pi r}\qquad p_r=-\rho\qquad p_\perp = \frac12\,p_r
}{eq:rind16}
Note that our effective fluid does not behave at all like a cold dark matter fluid modelled by dust ($p_r=p_\perp=0$). 
We stress that there is only one free parameter in our effective fluid description --- the Rindler acceleration $a$. 
The dominant energy condition \cite{waldgeneral} holds, provided that $a$ is negative. 
We found above that this is not the sign needed to model the galactic rotation curves and the Pioneer anomaly. 
Taking the fluid interpretation literally, we predict an unusual equation of state \eqref{eq:rind16} for dark matter that could be tested by combining gravitational lensing and rotation curve data, as explained in Ref.~\cite{Faber:2005xc}.
This could provide the simplest route to falsify our model.

\paragraph{Outlook}
Whether the anisotropic fluid \eqref{eq:rind15}, \eqref{eq:rind16} is ``real'' in any sense or just an effective way to incorporate IR effects (that either accumulate within GR or that follow from some modification of GR) is of considerable theoretical interest, but goes beyond the scope of the current Letter. 
Still, we mention an intriguing factoid: The most general spherically symmetric solution to $g^{\al\be}\nabla_\al\partial_\be R=0$ or to $g^{\al\be}\nabla_\al\nabla_\be R_{\mu\nu}=0$ or to the EOM arising in conformal Weyl gravity \cite{Mannheim:1988dj} is given by our solution \eqref{eq:rind1}, \eqref{eq:rind8}-\eqref{eq:rind9} [in the first case supplemented by a Reissner--Nordstr\"om term $Q^2/r^2$ in the Killing norm \eqref{eq:rind9}]. 
In all three cases the Rindler acceleration $a$ emerges as a constant of motion and therefore naturally depends on the specific system under consideration.

\acknowledgments

I thank Alan Guth, Roman Jackiw, Wolfgang Kummer, Philip Mannheim, Stacy McGaugh, Florian Preis, Dimitri Vassilevich and Richard Woodard for discussions.
I am grateful to Herbert Balasin and Tonguc Rador for independently pointing out an incorrect sign in Eq.~\eqref{eq:rind14}.

DG is supported by the START project Y435-N16 of the Austrian Science Foundation (FWF).


\end{document}